\documentstyle[12pt,psfig,draft,lscape,array,amssymb,longtable]{nature-ejb-new}

\def\simlt{\mathrel{\hbox{\rlap{\hbox{\lower4pt\hbox{$\sim$}}}\hbox{$<$}}}}
\def\simgt{\mathrel{\hbox{\rlap{\hbox{\lower4pt\hbox{$\sim$}}}\hbox{$>$}}}}

\def\ale{\mathrel{\hbox{\rlap{\hbox{\lower4pt\hbox{$\sim$}}}\hbox{$<$}}}}
\def\age{\mathrel{\hbox{\rlap{\hbox{\lower4pt\hbox{$\sim$}}}\hbox{$>$}}}}

\def\gtorder{\mathrel{\raise.3ex\hbox{$>$}\mkern-14mu
             \lower0.6ex\hbox{$\sim$}}}
\def\ltorder{\mathrel{\raise.3ex\hbox{$<$}\mkern-14mu
             \lower0.6ex\hbox{$\sim$}}}

\def\nodata{---}

\def\erancite#1{#1}
\def\citebraket#1{$^{[}$#1$^{]}$}
\def\citerange#1#2{$^{[}$#1$^{-}$#2$^{]}$}

\def\spose#1{\hbox to 0pt{#1\hss}}
\newcommand\lsim{\mathrel{\spose{\lower 3pt\hbox{$\mathchar"218$}}
     \raise 2.0pt\hbox{$\mathchar"13C$}}}
\newcommand\gsim{\mathrel{\spose{\lower 3pt\hbox{$\mathchar"218$}}
     \raise 2.0pt\hbox{$\mathchar"13E$}}}

\begin{document} 

\title{An outburst from a massive star 40 days before a supernova explosion}

\author{E. O. Ofek\affiliation[1]{Benoziyo Center for Astrophysics, Weizmann Institute of Science, 76100 Rehovot, Israel.},
        M. Sullivan\affiliation[2]{School of Physics and Astronomy, University of Southampton, Southampton, SO17 1BJ, UK.}$^{,}$\affiliation[3]{Department of Physics (Astrophysics), University of Oxford, Keble Road, Oxford, OX1 3RH, UK.},
        S. B. Cenko\affiliation[4]{Department of Astronomy, University of California, Berkeley, CA 94720-3411, USA.},
        M. M. Kasliwal\affiliation[5]{Observatories of the Carnegie Institution for Science, 813 Santa Barbara St., Pasadena, CA 91101, USA.},
        A. Gal-Yam\affiliationmark[1],
        S. R. Kulkarni\affiliation[6]{Division of Physics, Mathematics, and Astronomy, California Institute of Technology, Pasadena, CA 91125, USA.},
        I. Arcavi\affiliationmark[1],
        L. Bildsten\affiliation[7]{Kavli Institute for Theoretical Physics, Kohn Hall, University of California, Santa Barbara, CA 93106, USA.}$^{,}$\affiliation[8]{Department of Physics, Broida Hall, University of California, Santa Barbara, CA 93106, USA.},
        J. S. Bloom\affiliationmark[4]$^{,}$\affiliation[9]{Lawrence Berkeley National Laboratory, 1 Cyclotron Road, Berkeley, CA 94720, USA.},
        A. Horesh\affiliationmark[6],
        D. A. Howell\affiliationmark[8]$^{,}$\affiliation[10]{Las Cumbres Observatory Global Telescope Network, 6740 Cortona Dr., Suite 102, Goleta, CA 93117, USA.},
        A. V. Filippenko\affiliationmark[4],
        R. Laher\affiliation[11]{Spitzer Science Center, California Institute of Technology, M/S 314-6, Pasadena, CA 91125, USA.},
        D. Murray\affiliation[12]{Physics Department, University of Wisconsin--Milwaukee, Milwaukee, WI 53211, USA.},
        E. Nakar\affiliation[13]{School of Physics and Astronomy, Tel-Aviv University, Tel-Aviv, Israel.},
        P. E. Nugent\affiliationmark[9]$^{,}$\affiliationmark[4],
        J. M. Silverman\affiliationmark[4]$^{,}$\affiliation[14]{Department of Astronomy, University of Texas, Austin, TX 78712-0259, USA.},
        N. J. Shaviv\affiliation[15]{Racah Institute of Physics, The Hebrew University, 91904 Jerusalem, Israel.},
        J. Surace\affiliationmark[11],
        O. Yaron\affiliationmark[1]
}

\date{\today}{}
\headertitle{An outburst from a SN progenitor}
\mainauthor{Ofek et al.}

\summary{Various lines of evidence suggest that very massive stars
experience extreme mass-loss episodes
shortly before they explode as a 
supernova.\citerange{\cite{Dopita+1984}}{\cite{Ofek+2010}}
Interestingly, 
several models predict such pre-explosion 
outbursts.\citerange{\cite{Woosley+2007}}{\cite{Chevalier2012}}
Establishing a 
causal connection between these mass-loss episodes
and the final supernova explosion
will provide a novel way to study pre-supernova massive-star evolution.
Here we report on observations of a remarkable mass-loss event detected
40\,days prior to the explosion of the Type IIn supernova
SN\,2010mc (PTF\,10tel).
Our photometric and spectroscopic data
suggest that this event is a result of
an energetic outburst, radiating at least $6\times10^{47}$\,erg of energy,
and releasing about $10^{-2}$\,M$_{\odot}$ at
typical velocities of 2000\,km\,s$^{-1}$.
We show that the temporal proximity of the mass-loss outburst and the
supernova explosion implies a causal connection between them.
Moreover, we find that the outburst luminosity and velocity
are consistent with the
predictions of the wave-driven pulsation model\citebraket{\cite{Quataert+2012}},
and disfavour alternative suggestions\citebraket{\cite{Chevalier2012}}.}

\maketitle


Type IIn supernovae (SNe~IIn) are a diverse class of transient
events, spectroscopically defined by narrow and/or intermediate-width hydrogen emission lines,
up to a few thousand km\,s$^{-1}$ in 
width.\citerange{\cite{Schlegel1990}}{\cite{Kiewe+2012}}
These emission lines likely originate from optically thin
circumstellar matter excited by the SN shock and/or radiation field.
The spectra and light curves
of these SNe are interpreted as signatures of interaction
between the SN ejecta and mass that has been expelled
prior to the 
explosion.\citerange{\cite{Chugai+2003}}{\cite{Smith+2007}}
%
The best case in which the progenitor of a SN~IIn
has been detected prior to the SN
explosion\citebraket{\cite{Gal-Yam+2009}} was
likely a luminous blue variable (LBV) --
a class of objects which are known for their
vigorous eruptive mass-loss events\citebraket{\cite{Owocki2010}}.
%
Very recently, a massive star in the nearby galaxy NGC\,7259, SN\,2009ip,
had a series of 
outbursts\citerange{\cite{Smith+2010}}{\cite{Foley+2011}}
with a typical velocity of about 500\,km\,s$^{-1}$,
possibly followed by an ongoing
supernova event\citerange{\cite{Mauerhan+2012}}{\cite{Prieto+2012}}
exhibiting P-Cygni
lines with velocities of $\sim 10^{4}$\,km\,s$^{-1}$.
Another possibly related event is SN\,2006jc
that is further discussed in SI~\S{7}. 

\begin{figure}
\centerline{\psfig{file=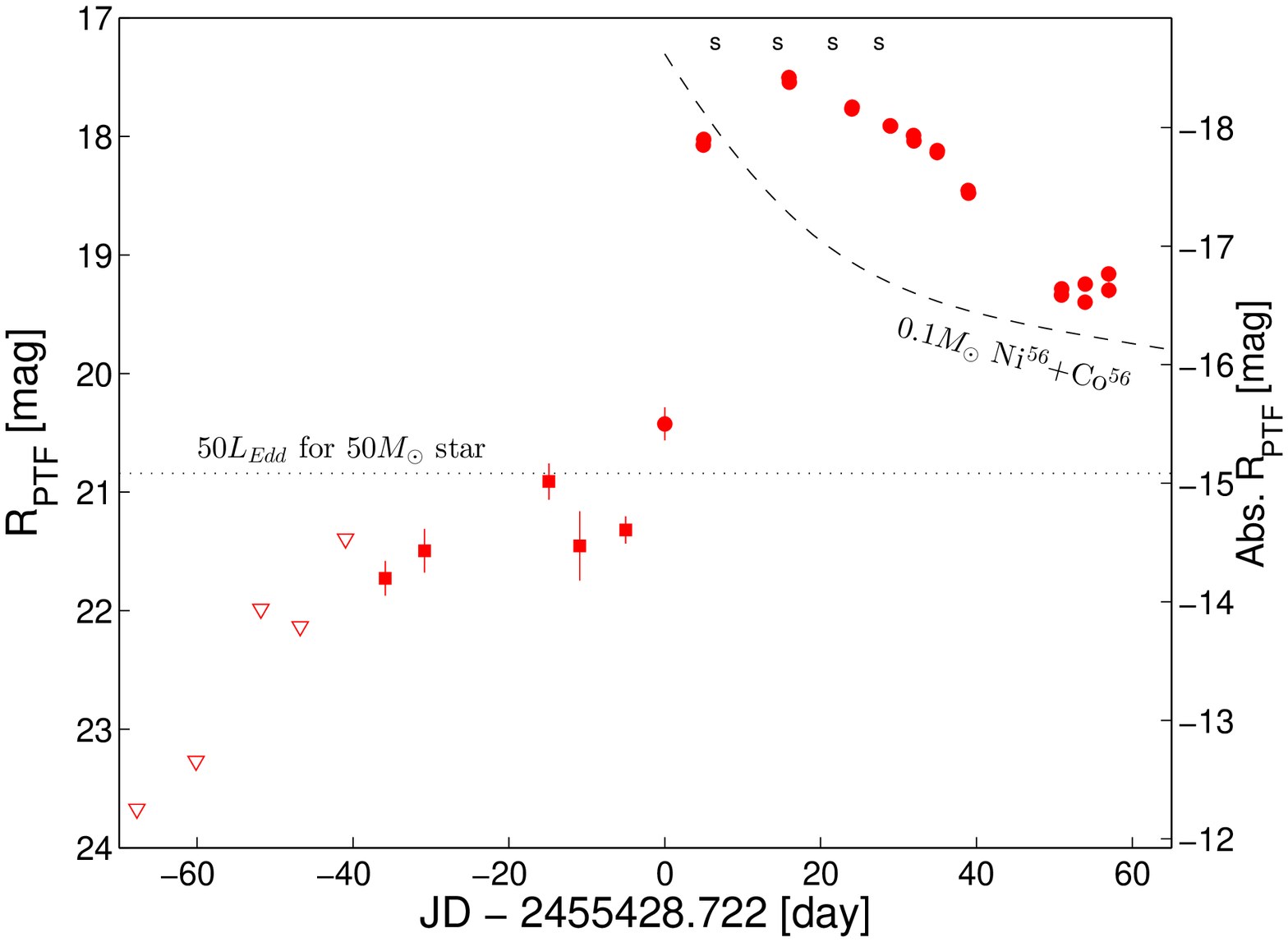,width=8in,angle=0}}
\caption[]{}
\label{fig:PTF10tel_PTF_LC}
\end{figure}

The Palomar Transient Factory\citerange{\cite{Law+2009}}{\cite{Rau+2009}}
discovered the Type IIn SN\,2010mc (PTF\,10tel)\citebraket{\cite{Ofek+2012c}}
in images obtained on 2010 Aug. 20.22 
(UTC dates are used throughout this paper).
It is located at $\alpha = 
17^{{\rm h}}21^{{\rm m}}30.^{{\rm s}}68$, $\delta = +48^{\circ}07'47.''4$ (J2000.0),
at redshift $z=0.035$ which corresponds
to a luminosity distance of 153\,Mpc.
Our ongoing search for precursor events
in pre-explosion images of nearby PTF SNe~IIn
revealed a positive detection at
the SN location in images taken prior 
to the SN discovery (Figure~\ref{fig:PTF10tel_PTF_LC}).
Here we measure time relative to 2010 Aug. 20.22,
which corresponds to the onset of the SN explosion
(main event; Fig.~\ref{fig:PTF10tel_PTF_LC}).
The initial bump emerged at day $-37$ relative
to the SN discovery date, and peaked at an absolute magnitude
of about $-15$ ($\sim 2.25\times10^{41}$\,erg\,s$^{-1}$)
in the $R_{{\rm PTF}}$ band.
The main SN explosion then brightened for two weeks
and peaked at an $R_{{\rm PTF}}$ absolute magnitude of $-18.4$
($\sim 5.2\times10^{42}$\,erg\,s$^{-1}$),
radiating a
total bolometric luminosity of
$\sim 3\times10^{49}$\,erg,
while the precursor bump radiated $\sim 6\times10^{47}$\,erg
(or more due to the unknown bolometric correction).

\begin{figure}
\centerline{\psfig{file=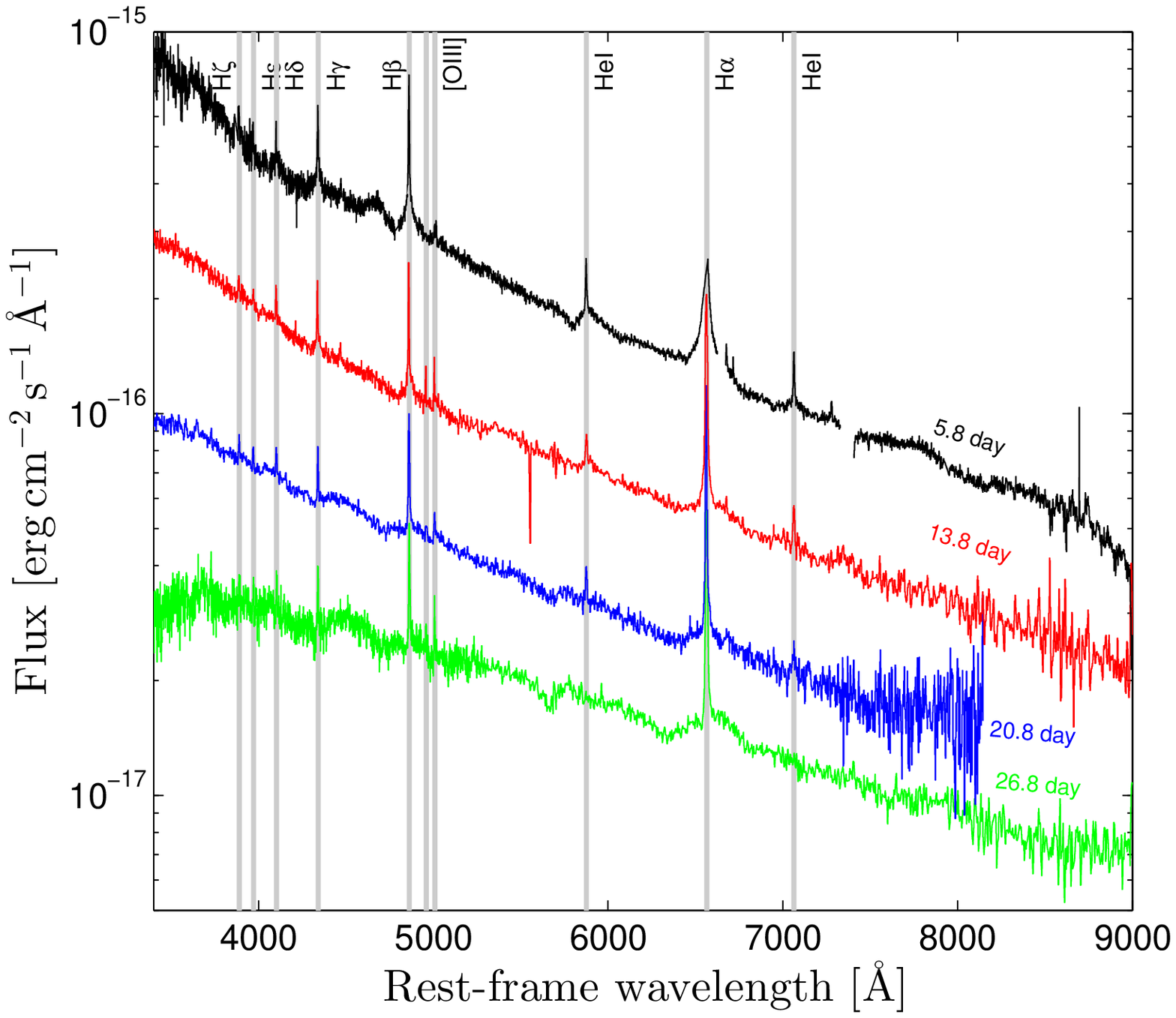,width=8in,angle=0}}
\caption[]{}
\label{fig:PTF10tel_spec}
\end{figure}

Spectra of the supernova, showing a blue continuum
with Balmer emission lines, are presented in Figure~\ref{fig:PTF10tel_spec}.
The continuum becomes redder with time,
and its slope corresponds to an effective temperature
of over 16,000\,K at day five and drops to about 8,000\,K
at day 27.
The H$\alpha$ line
has an initial width of $\sim3\times10^{3}$\,km\,s$^{-1}$
at day 6,
decreasing to $\sim10^{3}$\,km\,s$^{-1}$
at day 14.
A broad ($10^{4}$\,km\,s$^{-1}$) P-Cygni profile emerges
by day 27.
The spectra also show He\,I lines with decreasing strength,
presumably due to the drop in temperature.

The nature of the precursor bump is very intriguing
and can potentially tell us a great deal about the SN explosion
and the progenitor.
The only interpretation that is fully consistent
with the photometric and spectroscopic evidence
is that the first bump represents an outburst
from the SN progenitor about one month prior to explosion,
while the brighter bump is initiated by a full explosion
of the star a few weeks later.
Below we analyse this model in the context of the
photometric and spectroscopic data.
In SI \S{6}
we discuss some alternative models
and conclude that they are unlikely.

The mass ejected by the precursor burst can be estimated
in various independent ways.
By requiring that the precursor integrated bolometric luminosity,
$E_{{\rm bol,prec}}$,
is lower than the kinetic energy of the
precursor outburst (moving at velocity $v_{{\rm prec}}$)
which powers it,
we can set a lower limit on the mass ejected in the precursor outburst
$M_{{\rm prec}}\gtorder 2E_{{\rm bol,prec}} v_{{\rm prec}}^{-2} \approx
 1.5\times10^{-2} (v_{{\rm prec}}/2000\,{\rm km\,s}^{-1})^{-2}$\,M$_{\odot}$.
The outburst velocity is estimated from
the line widths of 1000--3000\,km\,s$^{-1}$,
seen in the early-time spectra of the SN.
As this mass was presumably ejected
over a period of about one month (i.e., the outburst duration),
the annual mass-loss rate is about 10 times higher.
A similar order of magnitude argument can be used to put an upper limit on
the mass in the precursor outburst.
If some of the SN bolometric energy, $E_{{\rm bol,SN}}$, is due to interaction
between the SN ejecta, moving at $v_{{\rm SN}}$, and the precursor shell,
and assuming high efficiency of conversion of the kinetic
energy to luminosity,
then $M_{{\rm prec}}\ltorder 2E_{{\rm bol,SN}}v_{{\rm SN}}^{-2}\approx 3\times10^{-2} (v_{{\rm SN}}/10^{4}\,{\rm km\,s}^{-1})^{-2}$\,M$_{\odot}$. 

Another method to estimate the mass in
the ejecta is based on the H$\alpha$ emission-line luminosity
and the radius of the photosphere,
which is determined from a black-body fit to the spectra
(SI Fig.~2).
This method, derived in SI \S{2},
suggests a mass-loss rate of $\gtorder 10^{-1}$\,M$_{\odot}$\,yr$^{-1}$,
or a total ejected mass of $\gtorder10^{-2}$\,M$_{\odot}$,
if we assume a month-long outburst with a mean wind velocity 
of 2000\,km\,s$^{-1}$.

Another independent method for estimating the mass
in the CSM is based on the rise time of the supernova.
Since photons diffuse through material between
the SN and the observer, the SN rise time gives us
an upper limit on the diffusion time scale and therefore
the total intervening mass.
Since the SN rise time was about one week,
this argument suggests that if the shell is spherically symmetric
its total mass cannot exceed about 0.4\,M$_{\odot}$
(see SI \S{5}).
Therefore, the kinetic-energy arguments,
the H$\alpha$ luminosity, the diffusion time scale,
and the X-ray limits (see SI \S{3})
are consistent with each other
and indicate that the total mass lost during the outburst
is of order $\sim 10^{-2}$\,M$_{\odot}$.

\begin{figure}
\centerline{\psfig{file=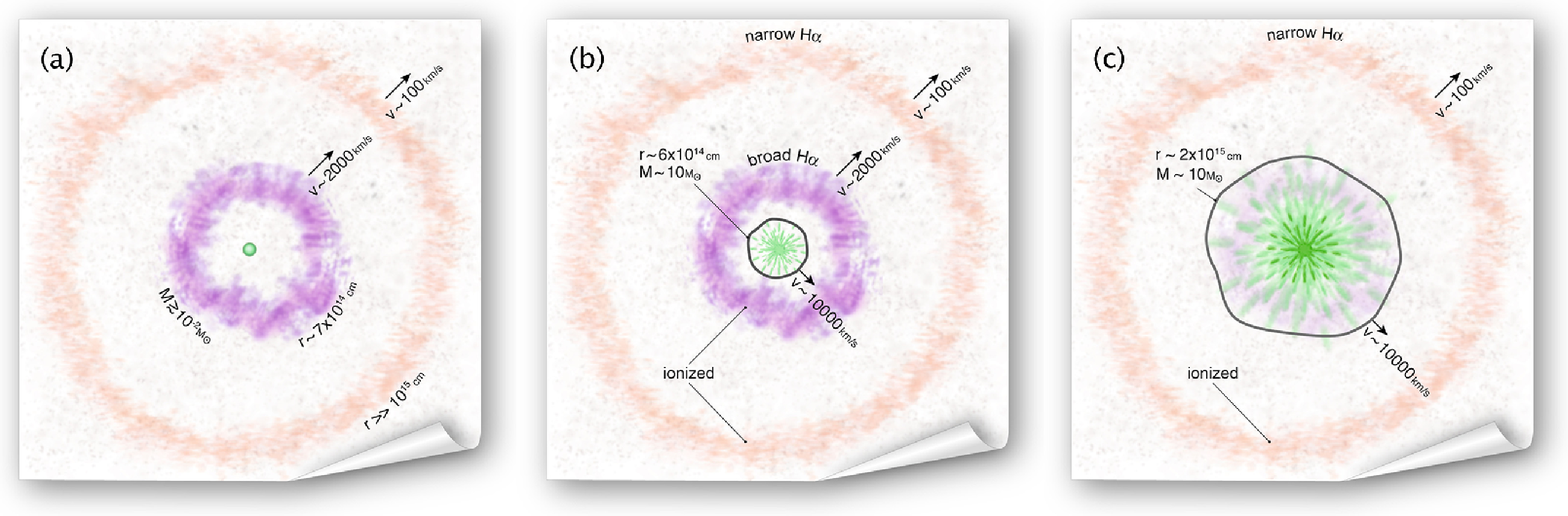,width=8in,angle=0}}
\caption[]{}
\label{fig:10tel_panels_top}
\end{figure}

Our model for the sequence of events is presented in
Figure~\ref{fig:10tel_panels_top}.
In a nutshell, this model suggests
that the precursor outburst ejected $\sim 10^{-2}$\,M$_{\odot}$
at a velocity of 2000\,km\,s$^{-1}$ about one month prior to the SN
explosion.
Shortly after the SN explosion, this ejected material
was engulfed by the SN ejecta.
At later times, after the optical depth decreases, we start
seeing indications for the high velocity of the SN ejecta.
We discuss some less likely alternative models in the SI\,\S{6}.

A surprising result is the short time
between the outburst and the explosion,
which is a tiny fraction ($\sim10^{-8}$) of the lifetime of a massive star.
Even if massive stars have multiple mass-loss episodes
(with mass loss $\gtorder 10^{-2}$\,M$_{\odot}$)
during their lifetime, the number of such episodes
cannot exceed $\sim5000$; otherwise, it will exceed a typical stellar mass.
A conservative estimate shows that
so far it might be possible to detect such an outburst 
in a sample of up to about 20 nearby 
SNe~IIn, which have deep pre-explosion observations
(see SI \S{7}).
Therefore, the probability of observing a random burst
one month prior to the explosion is $0.1\%$.
We conclude that such outbursts
are either causally related to, or at least two orders
of magnitude more common before, the final stellar explosion.

When considering where such an eruption could originate, one is led to
the stellar core, as the binding energy it can liberate by contraction
and nuclear reactions is sufficiently large. Because the actual luminosity
is most likely neutrino-dominated, the Kelvin-Helmholtz time scale for
this contraction can be short enough to explain the precursor's rapid rise
in luminosity. However, the energy liberated at the core must reach the
envelope, and it therefore requires 
a fast and efficient mechanism\citebraket{\cite{Quataert+2012}}.
There are several proposed mechanisms that
predict high mass-loss rates prior to the final stellar explosion.
Recently, it was suggested\citebraket{\cite{Quataert+2012}}
that in some massive
stars the super-Eddington fusion luminosities,
shortly prior to core collapse,
can drive convective motions that in turn excite
gravity waves that propagate toward the stellar surface.
The dissipation of these waves can unbind up to several
solar masses of the stellar envelope.
Alternatively, it was
suggested\citebraket{\cite{Chevalier2012}} that the mass loss
is driven by a common-envelope phase\citebraket{\cite{Taam+2010}}
due to the inspiral
of a neutron star into a giant companion core
(a so-called Thorne-\.{Z}ytkow object),
unbinding the companion envelope and setting up accretion
onto the neutron star that collapses into a black hole and
triggers a SN explosion shortly thereafter.
This model, however, predicts an outflow velocity that
is considerably lower than the one observed in
the precursor of SN\,2010mc, and is therefore disfavoured.
Another possible mechanism
is the pulsational pair instability SN\citebraket{\cite{Rakavy+1967}}.
However, current models
predict that the mass ejected in these events will likely be much larger
than the $10^{-2}$\,M$_{\odot}$ seen in SN\,2010mc
for the observed 40 day delay between the two
last explosions.\citebraket{\cite{Woosley+2007}}$^,$\citebraket{\cite{Quimby+2011}}

The velocity and energetics of the precursor of SN\,2010mc
are consistent with the predictions of the wave-driven
outburst mechanism\citebraket{\cite{Quataert+2012}}.
However, a more detailed theoretical investigation
is required in order to test that this model is
consistent with all the observational evidence.
Moreover, in SI \S{8},
given the observed precursor luminosity,
we theoretically derive the mass loss
to be $\sim0.05$\,M$_{\odot}$,
in excellent agreement with our mass-loss estimates.
Furthermore, we note that
the velocity of the precursor ejecta
is higher than predicted by other models.\citebraket{\cite{Chevalier2012}}
Finally, we note that the mass-loss
velocity in SN\,2010mc is considerably higher
than the one observed in SN\,2009ip,\citebraket{\cite{Mauerhan+2012}}
although the mass-loss rate was similar.\citebraket{\cite{Ofek+2012d}}

\bibliographystyle{nature-pap}

\noindent 
{\bf \large Supplementary Information} is linked to the online version of the
paper at www.nature.com/nature.

\begin{acknowledge}
We thank Eliot Quataert and Matteo Cantiello for helpful discussions.
The VLA is operated by the National Radio
Astronomy Observatory, a facility of the U.S. National Science Foundation (NSF)
operated under cooperative agreement by Associated Universities, Inc.
This paper is based on observations obtained with the
Samuel Oschin Telescope as part of the Palomar Transient Factory
project.
We are grateful for the assistance of the staffs at the various 
observatories where data were obtained.
The authors acknowledge support from
the Arye Dissentshik career development chair,
the Helen Kimmel Center for Planetary Science,
the Israeli Ministry of Science,
the Royal Society,
the NSF,
the Israeli Science Foundation,
the German-Israeli Foundation,
ERC,
the U.S. Department of Energy,
Gary \& Cynthia Bengier, the Richard \& Rhoda Goldman Fund, the
Christopher R. Redlich Fund, and the TABASGO Foundation.

\end{acknowledge}

\bigskip
\noindent

{\bf Author contribution}
E.~O.~Ofek initiated the search for precursor outbursts and wrote the paper,
M.~Sullivan wrote the careful image-subtraction pipeline,
S.~B.~Cenko reduced the {\it Swift}-UVOT observations,
M.~M.~Kasliwal reduced the P60 observations,
A.~Gal-Yam helped with the the spectroscopic analysis,
L.~Bildsten, E.~Nakar, and N.~J.~Shaviv contributed to the theoretical interpretation,
A.~V.~Filippenko assisted with the spectroscopy and edited the paper,
D.~A.~Howell, D.~Murray, and J.~M.~Silverman conducted the spectroscopic observations or reductions,
O.~Yaron helped write the paper, and
S.~R.~Kulkarni, I.~Arcavi, J.~S.~Bloom, R.~Laher, P.~E.~Nugent, and J.~Surace built the PTF
hardware and software infrastructure. 

\bigskip
\noindent

Reprints and permissions information is available at
npg.nature.com/reprintsandpermissions.  Correspondence should be
addressed to E.~O.~Ofek (e-mail: eran@astro.caltech.edu).

\clearpage

\bigskip \noindent {\bf Figure~1:}
{\bf The light curve of SN\,2010mc as obtained with the Palomar 48-inch telescope.}
The red circles are
based on individual images,
the squares are based on coadded images,
while the empty triangles represent the 3$\sigma$ upper limits
derived from coadded images.
The error bars represent the 1$\sigma$ errors.
The object magnitudes are given in the PTF magnitude 
system.\citebraket{\cite{Ofek+2012a}}$^,$\citebraket{\cite{Ofek+2012b}}
Other datasets, including the Palomar 60-inch and {\it Swift}-UVOT,
are listed in the SI (Table~3).
All the photometric and spectroscopic data are available
via the WISeREP website\citebraket{\cite{Yaron+2012}}.
The dashed line shows the expected luminosity from
the radioactive decay of an ejected mass of
0.1\,M$_{\odot}$ of Ni$^{56}$.
Assuming that at late times the optical depth is sufficiently large
to convert the radioactive energy to optical luminosity,
but not too large so it will go into $PdV$ work,
this line represents an upper limit on
the total amount of Ni$^{56}$ in the ejecta;
it was set to coincide with the latest observation
of the SN at JD $\approx$ 2,455,758 (see Table~3).
The dotted line represents a bolometric luminosity equal
to 50 times the Eddington luminosity for a 50\,M$_{\odot}$ star
(order of magnitude
estimate of the mass of the progenitor assuming it is a massive star).
The right edges of the ``S'' symbols above the light curve indicate the
epochs at which we obtained spectra (see Fig.~\ref{fig:PTF10tel_spec}).
A full version of this light curve, including the late-time
observations, is shown in the SI.

\bigskip \noindent {\bf Figure~2:}
{\bf Spectra of SN\,2010mc, 
showing prominent Balmer emission lines.}
The early-time spectra are well fitted by a narrow component
($\sim10^{2}$\,km\,s$^{-1}$; the fit includes the
instrumental line broadening)
and a broad component extending
to $\sim3\times10^{3}$\,km\,s$^{-1}$ at the first epoch,
and decreasing to $10^{3}$\,km\,s$^{-1}$ at the second epoch.
At later epochs the H$\alpha$ line develops a P-Cygni profile
(similar to that of normal SNe~II)
with a velocity difference between the absorption bottom 
and the emission peak
of $\sim10^{4}$\,km\,s$^{-1}$.
At the first epochs we also detect He\,I
lines, some of which are marked on the plot.
They become weaker (see SI Figure~2)
at later epochs,
presumably due to the decrease in the effective temperature
(see SI Figure~2).
This, as well as the absence of He\,II lines,
indicates that our temperature
estimate (SI Figure~2)
is reasonable.

\bigskip \noindent {\bf Figure~3:}
{\bf Qualitative sketch of the proposed model for SN\,2010mc.}
{\it Panel a}:
At day $\sim0$, an inner shell (purple)
with a mass of $\sim10^{-2}$\,M$_{\odot}$,
ejected about one month earlier, during the precursor outburst, and
moving at about 2000\,km\,s$^{-1}$, is located
at a radius of $\sim7\times10^{14}$\,cm.
An outer shell (orange), found at a large radius and moving at a few 
hundreds km\,s$^{-1}$ (up to $10^{3}$\,km\,s$^{-1}$),
was ejected at earlier times.
This indicates that the progenitor likely had multiple mass-loss episodes
in the past tens to hundreds of years prior to the explosion.
{\it Panel b}:
At day $\sim5$, the SN shock front
(grey line) moving at $\sim10^{4}$\,km\,s$^{-1}$
is ionising the inner and outer shells
which produce the broad and narrow H$\alpha$ emission
seen in the early-time spectra.
{\it Panel c}:
At day $\sim20$, the SN shock engulfs the
inner shell, and the intermediate width ($\sim2000$\,km\,s$^{-1}$) component
of the H$\alpha$ line disappears.
Instead we detect a 1000\,km\,s$^{-1}$ line,
presumably due to material ejected during previous, but likely recent,
mass-loss episodes and that is found at
larger distances from the SN.
We note that inspection of the SN light curve shows that around day 50
there is  an indication for a possible rebrightening,
perhaps resulting from the SN ejecta colliding with
such additional material ejected at earlier times.
At day $\sim20$, the photospheric temperature
decreases and it becomes optically thinner,
and therefore we begin seeing an H$\alpha$ P-Cygni profile
with a velocity of $\sim 10^{4}$\,km\,s$^{-1}$.
This line become even stronger on day 27.
This reflects the unshocked ejecta below the interaction zone.

\clearpage

\clearpage

\clearpage

\begin{center}
{\Large SUPPLEMENTARY INFORMATION} \\
\end{center}

\headertitle{Suppl. Info.}
\mainauthor{Ofek et al.}

\section{Observations}
\label{sec:obs}

\subsection{Photometry and spectroscopy}
\label{sec:phot}

On 2010 Aug. 25, the Palomar Transient Factory
(PTF; \erancite{Law et al. 2009}; \erancite{Rau et al. 2009}) discovered
with an autonomous machine-learning framework (Bloom et al. 2011)
the Type IIn SN\,2010mc (PTF\,10tel; \erancite{Ofek et al. 2012c}) in images
obtained on 2010 Aug. 20.22 UT.
Further inspection of the images generated by
the automatic image-subtraction pipeline revealed that the object
was detected even on Aug. 20.22 (i.e., the discovery date).

We reanalysed the images using a careful image-subtraction
pipeline. The improved reduction clearly shows
a variable source at the location of the SN
prior to the discovery date.
Table~\ref{tab:phot} presents the photometric measurements
of SN\,2010mc, including observations made with 
the Oschin Schmidt 48-inch telescope,
the Palomar 60-inch telescope (\erancite{Cenko et al. 2006}),
and the {\it Swift}-UVOT (\erancite{Gehrels et al. 2004}).
We note that the ground-based observations were
processed using image subtraction.

\begin{figure}
\centerline{\psfig{file=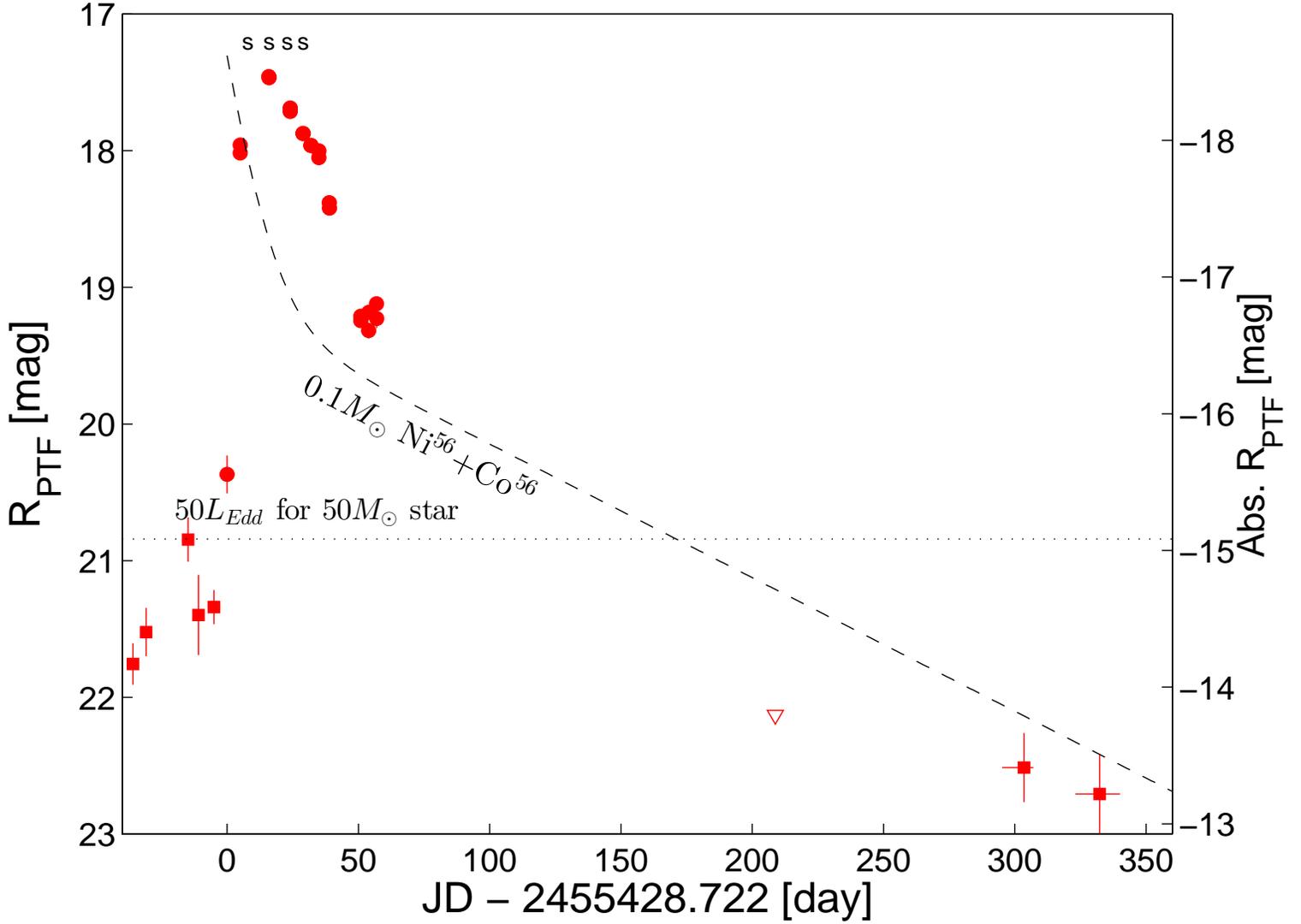,width=8in,angle=0}}
\caption[]{\small 
The light curve of SN\,2010mc (PTF\,10tel) as obtained with
the Palomar 48-inch telescope, including late-time observations.
See main text Figure~1 for details.
}
\label{fig:PTF10tel_PTF_LC_Full}
\end{figure}

A version of the light curve showing the late-time observations
of the SN is presented in SI Figure~\ref{fig:PTF10tel_PTF_LC_Full}.

\clearpage

\begin{center}
\begin{longtable}{lllllllll}
\hline\hline
Telescope & Band  & Time  & $\Delta_{-}$ & $\Delta_{+}$ & N & AB Mag & Err & Mag limit \\
         &       & (day)   & (day)         & (day)          &   & (mag) & (mag) & (mag) \\
\hline
       PTF&   R&    -0.0000&     \nodata&     \nodata&   1&  20.42&  0.14&  21.46\\[-3.3mm]
       PTF&   R&     4.9600&     \nodata&     \nodata&   1&  18.07&  0.02&  21.34\\[-3.3mm]
       PTF&   R&     5.0030&     \nodata&     \nodata&   1&  18.02&  0.02&  21.17\\[-3.3mm]
       PTF&   R&    15.9350&     \nodata&     \nodata&   1&  17.50&  0.01&  21.05\\[-3.3mm]
       PTF&   R&    15.9820&     \nodata&     \nodata&   1&  17.54&  0.01&  21.22\\[-3.3mm]
       PTF&   R&    24.0140&     \nodata&     \nodata&   1&  17.77&  0.01&  21.32\\[-3.3mm]
       PTF&   R&    24.0580&     \nodata&     \nodata&   1&  17.75&  0.02&  20.93\\[-3.3mm]
       PTF&   R&    28.9200&     \nodata&     \nodata&   1&  17.91&  0.01&  21.54\\[-3.3mm]
       PTF&   R&    28.9640&     \nodata&     \nodata&   1&  17.91&  0.01&  21.47\\[-3.3mm]
       PTF&   R&    31.9190&     \nodata&     \nodata&   1&  17.99&  0.02&  21.42\\[-3.3mm]
       PTF&   R&    31.9770&     \nodata&     \nodata&   1&  18.04&  0.02&  21.26\\[-3.3mm]
       PTF&   R&    34.9320&     \nodata&     \nodata&   1&  18.14&  0.02&  21.43\\[-3.3mm]
       PTF&   R&    34.9760&     \nodata&     \nodata&   1&  18.12&  0.04&  20.59\\[-3.3mm]
       PTF&   R&    38.9240&     \nodata&     \nodata&   1&  18.45&  0.03&  21.20\\[-3.3mm]
       PTF&   R&    38.9670&     \nodata&     \nodata&   1&  18.48&  0.04&  21.01\\[-3.3mm]
       PTF&   R&    50.8910&     \nodata&     \nodata&   1&  19.34&  0.03&  22.07\\[-3.3mm]
       PTF&   R&    50.9350&     \nodata&     \nodata&   1&  19.28&  0.03&  21.84\\[-3.3mm]
       PTF&   R&    53.9120&     \nodata&     \nodata&   1&  19.40&  0.06&  21.39\\[-3.3mm]
       PTF&   R&    53.9570&     \nodata&     \nodata&   1&  19.25&  0.05&  21.40\\[-3.3mm]
       PTF&   R&    56.9350&     \nodata&     \nodata&   1&  19.16&  0.04&  21.58\\[-3.3mm]
       PTF&   R&    56.9790&     \nodata&     \nodata&   1&  19.30&  0.07&  21.12\\[-3.3mm]
       PTF&   R&   -35.8710&     \nodata&     \nodata&   1&  21.74&  0.21&  22.32\\[-3.3mm]
       PTF&   R&   -35.8260&     \nodata&     \nodata&   1&  21.71&  0.20&  22.35\\[-3.3mm]
       PTF&   R&   -30.8300&     \nodata&     \nodata&   1&  21.49&  0.18&  22.23\\[-3.3mm]
       PTF&   R&   -14.8740&     \nodata&     \nodata&   1&  20.98&  0.18&  21.72\\[-3.3mm]
       PTF&   R&    -5.0370&     \nodata&     \nodata&   1&  21.45&  0.24&  21.90\\[-3.3mm]
       PTF&   R&    -4.9930&     \nodata&     \nodata&   1&  21.27&  0.13&  22.38\\[-3.3mm]
       PTF&   R&   -35.8485&  0.0225&  0.0225&   2&  21.73&  0.15&  22.71\\[-3.3mm]
       PTF&   R&   -30.8300&  0.0000&  0.0000&   1&  21.49&  0.18&  22.23\\[-3.3mm]
       PTF&   R&   -14.8520&  0.0220&  0.0220&   2&  20.91&  0.15&  21.85\\[-3.3mm]
       PTF&   R&   -10.9125&  0.0225&  0.0225&   2&  21.45&  0.29&  21.69\\[-3.3mm]
       PTF&   R&    -5.0150&  0.0220&  0.0220&   2&  21.32&  0.11&  22.57\\[-3.3mm]
       PTF&   R&   303.5230&  8.3140&  3.5980&   5&  22.44&  0.23&  22.93\\[-3.3mm]
       PTF&   R&   332.5799&  9.5219&  7.5501&  10&  22.64&  0.28&  22.92\\[-3.3mm]
       PTF&   R&  -125.7380& 21.9790& 11.0110&   3&    \nodata&   \nodata&  22.43\\[-3.3mm]
       PTF&   R&   -96.3635& 12.5635& 13.5765&  11&    \nodata&   \nodata&  23.48\\[-3.3mm]
       PTF&   R&   -70.3692&  7.4648&  8.5832&  12&    \nodata&   \nodata&  23.65\\[-3.3mm]
       PTF&   R&   -67.7174& 10.1166& 15.9324&  14&    \nodata&   \nodata&  23.67\\[-3.3mm]
       PTF&   R&   -60.1392&  6.6858&  8.3542&   6&    \nodata&   \nodata&  23.27\\[-3.3mm]
       PTF&   R&   -51.8065&  0.0215&  0.0215&   2&    \nodata&   \nodata&  21.99\\[-3.3mm]
       PTF&   R&   -51.8065&  0.0215&  0.0215&   2&    \nodata&   \nodata&  21.99\\[-3.3mm]
       PTF&   R&   -46.7715&  0.0225&  0.0225&   2&    \nodata&   \nodata&  22.13\\[-3.3mm]
       PTF&   R&   -40.9400&  0.0000&  0.0000&   1&    \nodata&   \nodata&  21.39\\[-3.3mm]
       PTF&   R&   208.7957&  1.5128&  1.5122&   4&    \nodata&   \nodata&  22.19\\[-3.3mm]
       PTF&   R&   355.6692&  4.6512&  6.3248&   6&    \nodata&   \nodata&  23.16\\[-3.3mm]
       PTF&   R&   387.0123&  4.0043&  7.9167&   3&    \nodata&   \nodata&  22.48\\[-3.3mm]
       PTF&   R&   419.9272&  4.0323&  4.0317&   4&    \nodata&   \nodata&  22.43\\[-3.3mm]
       P60&   g&     5.9400&     \nodata&     \nodata&   1&  17.96&  0.18&    \nodata\\[-3.3mm]
       P60&   g&     7.9630&     \nodata&     \nodata&   1&  17.89&  0.17&    \nodata\\[-3.3mm]
       P60&   g&    10.0990&     \nodata&     \nodata&   1&  17.78&  0.17&    \nodata\\[-3.3mm]
       P60&   g&    10.1030&     \nodata&     \nodata&   1&  17.78&  0.15&    \nodata\\[-3.3mm]
       P60&   g&    14.0070&     \nodata&     \nodata&   1&  17.70&  0.20&    \nodata\\[-3.3mm]
       P60&   g&    15.0140&     \nodata&     \nodata&   1&  17.62&  0.17&    \nodata\\[-3.3mm]
       P60&   g&    16.0330&     \nodata&     \nodata&   1&  17.61&  0.21&    \nodata\\[-3.3mm]
       P60&   g&    17.0640&     \nodata&     \nodata&   1&  17.59&  0.21&    \nodata\\[-3.3mm]
       P60&   g&    18.0750&     \nodata&     \nodata&   1&  17.60&  0.23&    \nodata\\[-3.3mm]
       P60&   g&    19.0210&     \nodata&     \nodata&   1&  17.74&  0.17&    \nodata\\[-3.3mm]
       P60&   g&    20.0700&     \nodata&     \nodata&   1&  17.75&  0.19&    \nodata\\[-3.3mm]
       P60&   g&    21.9430&     \nodata&     \nodata&   1&  17.69&  0.22&    \nodata\\[-3.3mm]
       P60&   g&    23.0360&     \nodata&     \nodata&   1&  17.86&  0.22&    \nodata\\[-3.3mm]
       P60&   g&    32.0150&     \nodata&     \nodata&   1&  18.30&  0.22&    \nodata\\[-3.3mm]
       P60&   g&    34.9860&     \nodata&     \nodata&   1&  18.56&  0.22&    \nodata\\[-3.3mm]
       P60&   g&    40.9780&     \nodata&     \nodata&   1&  19.04&  0.48&    \nodata\\[-3.3mm]
       P60&   g&    49.9340&     \nodata&     \nodata&   1&  19.91&  0.49&    \nodata\\[-3.3mm]
       P60&   g&    50.9070&     \nodata&     \nodata&   1&  19.85&  0.47&    \nodata\\[-3.3mm]
       P60&   g&    51.9660&     \nodata&     \nodata&   1&  19.62&  0.67&    \nodata\\[-3.3mm]
       P60&   g&    52.9400&     \nodata&     \nodata&   1&  19.59&  0.62&    \nodata\\[-3.3mm]
       P60&   g&    53.9140&     \nodata&     \nodata&   1&  19.95&  0.53&    \nodata\\[-3.3mm]
       P60&   g&    55.9820&     \nodata&     \nodata&   1&  19.80&  0.38&    \nodata\\[-3.3mm]
       P60&   g&    56.9000&     \nodata&     \nodata&   1&  19.68&  0.52&    \nodata\\[-3.3mm]
       P60&   g&    69.9530&     \nodata&     \nodata&   1&  20.42&  0.37&    \nodata\\[-3.3mm]
       P60&   g&    72.8890&     \nodata&     \nodata&   1&  20.23&  0.61&    \nodata\\[-3.3mm]
       P60&   r&     7.9610&     \nodata&     \nodata&   1&  18.08&  0.14&    \nodata\\[-3.3mm]
       P60&   r&    10.0960&     \nodata&     \nodata&   1&  17.97&  0.17&    \nodata\\[-3.3mm]
       P60&   r&    10.1010&     \nodata&     \nodata&   1&  18.00&  0.17&    \nodata\\[-3.3mm]
       P60&   r&    14.0060&     \nodata&     \nodata&   1&  17.93&  0.15&    \nodata\\[-3.3mm]
       P60&   r&    15.0130&     \nodata&     \nodata&   1&  17.78&  0.15&    \nodata\\[-3.3mm]
       P60&   r&    16.0320&     \nodata&     \nodata&   1&  17.84&  0.16&    \nodata\\[-3.3mm]
       P60&   r&    17.0630&     \nodata&     \nodata&   1&  17.80&  0.14&    \nodata\\[-3.3mm]
       P60&   r&    18.0740&     \nodata&     \nodata&   1&  17.84&  0.14&    \nodata\\[-3.3mm]
       P60&   r&    19.0200&     \nodata&     \nodata&   1&  17.88&  0.14&    \nodata\\[-3.3mm]
       P60&   r&    20.0690&     \nodata&     \nodata&   1&  17.82&  0.15&    \nodata\\[-3.3mm]
       P60&   r&    21.9420&     \nodata&     \nodata&   1&  17.77&  0.15&    \nodata\\[-3.3mm]
       P60&   r&    23.0350&     \nodata&     \nodata&   1&  17.88&  0.15&    \nodata\\[-3.3mm]
       P60&   r&    32.0140&     \nodata&     \nodata&   1&  18.33&  0.14&    \nodata\\[-3.3mm]
       P60&   r&    34.0740&     \nodata&     \nodata&   1&  18.39&  0.16&    \nodata\\[-3.3mm]
       P60&   r&    49.9320&     \nodata&     \nodata&   1&  19.57&  0.15&    \nodata\\[-3.3mm]
       P60&   r&    51.9640&     \nodata&     \nodata&   1&  19.49&  0.26&    \nodata\\[-3.3mm]
       P60&   r&    53.9130&     \nodata&     \nodata&   1&  19.56&  0.21&    \nodata\\[-3.3mm]
       P60&   r&    55.9800&     \nodata&     \nodata&   1&  19.34&  0.27&    \nodata\\[-3.3mm]
       P60&   r&    72.8870&     \nodata&     \nodata&   1&  20.08&  0.21&    \nodata\\[-3.3mm]
       P60&   i&     7.9590&     \nodata&     \nodata&   1&  18.33&  0.16&    \nodata\\[-3.3mm]
       P60&   i&    10.0950&     \nodata&     \nodata&   1&  18.07&  0.19&    \nodata\\[-3.3mm]
       P60&   i&    10.1000&     \nodata&     \nodata&   1&  18.28&  0.16&    \nodata\\[-3.3mm]
       P60&   i&    14.0010&     \nodata&     \nodata&   1&  18.24&  0.17&    \nodata\\[-3.3mm]
       P60&   i&    15.0120&     \nodata&     \nodata&   1&  18.02&  0.16&    \nodata\\[-3.3mm]
       P60&   i&    16.0310&     \nodata&     \nodata&   1&  18.08&  0.14&    \nodata\\[-3.3mm]
       P60&   i&    17.0620&     \nodata&     \nodata&   1&  18.00&  0.16&    \nodata\\[-3.3mm]
       P60&   i&    18.0730&     \nodata&     \nodata&   1&  17.98&  0.15&    \nodata\\[-3.3mm]
       P60&   i&    19.0190&     \nodata&     \nodata&   1&  18.10&  0.18&    \nodata\\[-3.3mm]
       P60&   i&    20.0580&     \nodata&     \nodata&   1&  18.01&  0.19&    \nodata\\[-3.3mm]
       P60&   i&    21.9410&     \nodata&     \nodata&   1&  17.88&  0.18&    \nodata\\[-3.3mm]
       P60&   i&    23.0230&     \nodata&     \nodata&   1&  17.91&  0.16&    \nodata\\[-3.3mm]
       P60&   i&    32.0130&     \nodata&     \nodata&   1&  18.52&  0.11&    \nodata\\[-3.3mm]
       P60&   i&    34.0720&     \nodata&     \nodata&   1&  18.52&  0.14&    \nodata\\[-3.3mm]
       P60&   i&    49.9310&     \nodata&     \nodata&   1&  19.59&  0.13&    \nodata\\[-3.3mm]
       P60&   i&    53.9110&     \nodata&     \nodata&   1&  19.56&  0.15&    \nodata\\[-3.3mm]
       P60&   i&    55.9790&     \nodata&     \nodata&   1&  19.28&  0.14&    \nodata\\[-3.3mm]
       P60&   i&    68.9290&     \nodata&     \nodata&   1&  19.25&  0.15&    \nodata\\[-3.3mm]
       P60&   i&    74.9270&     \nodata&     \nodata&   1&  19.20&  0.14&    \nodata\\[-3.3mm]
{\it Swift}/UVOT& UVW1 & \nodata & \nodata & 1 & 6.753  & 17.59 & 0.05 & \nodata \\[-3.3mm]
{\it Swift}/UVOT& UVW2 & \nodata & \nodata & 1 & 6.738  & 17.67 & 0.03 & \nodata \\[-3.3mm]
{\it Swift}/UVOT& UVW1 & \nodata & \nodata & 1 & 11.174 & 17.69 & 0.04 & \nodata \\[-3.3mm]
{\it Swift}/UVOT& UVW2 & \nodata & \nodata & 1 & 11.167 & 17.87 & 0.04 & \nodata \\[-3.3mm]
{\it Swift}/UVOT& UVW1 & \nodata & \nodata & 1 & 691.08 & 21.05 & 0.10 & \nodata \\[-3.3mm]
{\it Swift}/UVOT& U    & \nodata & \nodata & 1 & 691.88 & 20.54 & 0.08 & \nodata \\
\hline
\caption[Table~1]{SN\,2010mc (PTF\,10tel) photometric measurements.
Time is measured relative to JD = 2,455,428.722\,day.
$\Delta_{-}$ and $\Delta_{+}$ are the lower and upper time
range in which the photometry was obtained.
These columns are indicated in cases where we coadded
data from several images.
$N$ is the number of images.
PTF magnitudes are given
in the PTF magnitude system
(Ofek et al. 2012a; Ofek et al. 2012b).
The {\it Swift}/UVOT measurements
represents the aperture magnitude at the position of the SN and include the host-galaxy
light. The two last UVOT measurements were taken at late times; thus, they can be regarded as the magnitudes of the host galaxy. Based on them, we conclude that in order to correct the early-time measurements for the host-galaxy contamination, about 0.05\,mag should be added to the SN magnitudes.} 
\label{tab:phot} \\
\end{longtable}
\end{center}

\clearpage

\begin{figure}
\centerline{\psfig{file=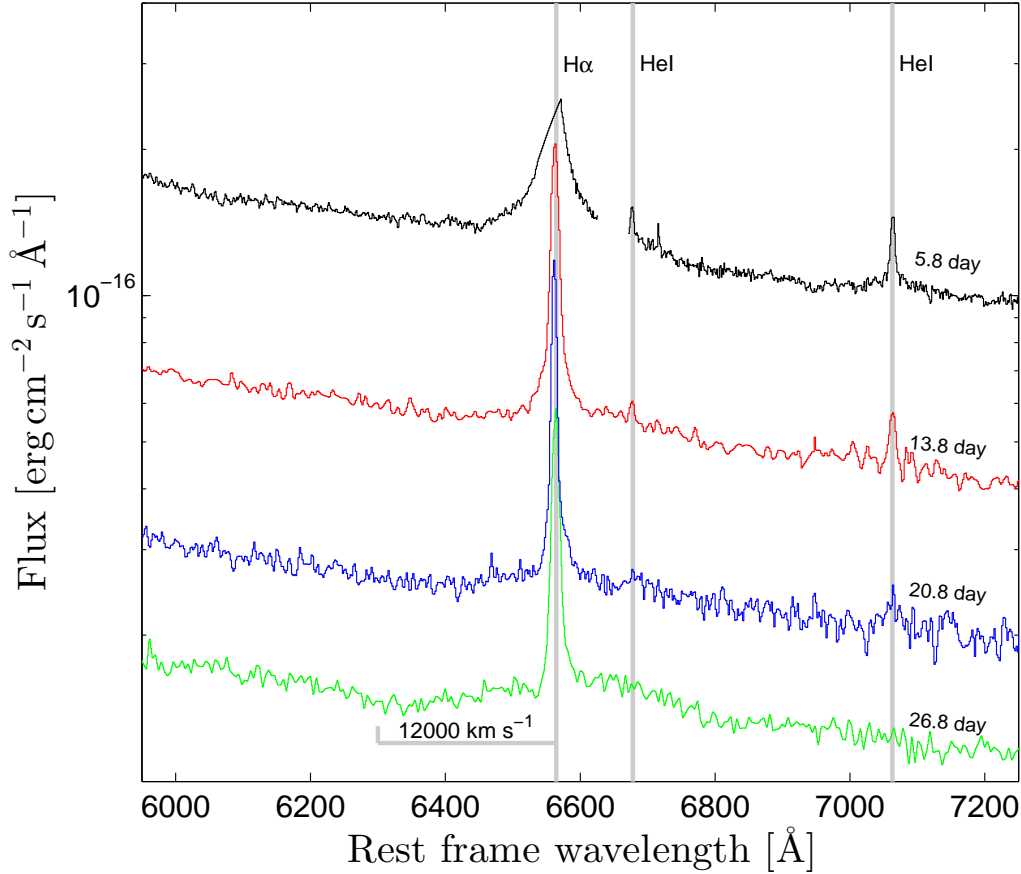,width=6in,angle=0}}
\caption[]{\small 
Close-up view of the H$\alpha$ line in the spectra of SN\,2010mc (PTF\,10tel).
The best-fit width and total flux of a two-component model
at each epoch are listed in Table~\ref{tab:spec}.
The width of the H$\alpha$ emission is decreasing
significantly after the first epoch,
from 3400\,km\,s$^{-1}$ to $\ltorder 1000$\,km\,s$^{-1}$.
In the last epoch a P-Cygni profile develops,
with a velocity difference between the peak of the emission
and the bottom of the absorption of over $10^{4}$\,km\,s$^{-1}$.
The velocity of the H$\alpha$ absorption component in the P-Cygni 
feature is marked with a 12,000 km\,s$^{-1}$ label.
The P-Cygni profile is better seen in the stretch of
Figure~2 (main text).
Also visible is the decrease in strength of the He\,I lines,
presumably due to the decrease in temperature
(see SI Figure~\ref{fig:PTF10tel_radtemp}).
}
\label{fig:PTF10tel_spec_Halpha} 
\end{figure}

In addition, we obtained spectra
of SN\,2010mc on four epochs, as listed in Table~\ref{tab:spec}
and presented in Figure~2 of the main paper and SI Figure~\ref{fig:PTF10tel_spec_Halpha}.
\begin{table}
\begin{center}
\begin{tabular}{llllll}
\hline\hline
UT Date & Telescope & Instrument & H$\alpha$ flux & broad H$\alpha$ vel. & narrow H$\alpha$ vel.   \\
     &           &            & (erg\,cm$^{-2}$\,s$^{-1}$) & (km\,s$^{-1}$) & (km\,s$^{-1}$)             \\
\hline
2010 Aug. 26 & Gemini & GMOS            &$9.2\times10^{-15}$&  3400& 800 \\
2010 Sep. 03 & Lick   & Kast            &$1.2\times10^{-14}$&  1100& 160 \\
2010 Sep. 10 & KPNO   & RC spectrograph &$9.2\times10^{-15}$&   670& 110 \\
2010 Sep. 16 & Lick   & Kast            &$1.4\times10^{-14}$&  4000& 230 \\
\hline
\end{tabular}
\caption[Figure~1]{\small Log of spectroscopic observations of SN\,2010mc (PTF\,10tel) and properties of the H$\alpha$ emission line. The fluxes and velocities were measured by fitting two-component Gaussians which were convolved
with the instrumental broadening as estimated from atmospheric emission lines.}   
\label{tab:spec} 
\end{center}
\end{table}
In order to measure the line fluxes in a consistent way,
we first calibrated the flux of each spectrum
such that the synthetic $R_{{\rm PTF}}$ magnitude
derived from the spectrum was identical
to the $R_{{\rm PTF}}$ magnitude, interpolated
to the epoch of each spectrum, from the PTF magnitudes
listed in Table~\ref{tab:phot}.
All of the spectra are available online from the
WISeREP website\footnote{http://www.weizmann.ac.il/astrophysics/wiserep/; 
Weizmann\ Interactive Supernova (data) REPository.}
(Yaron \& Gal-Yam 2012).
Each spectrum was smoothed using a median filter
and a black-body curve was fitted.
The derived effective temperatures and radii are
presented in SI Figure~\ref{fig:PTF10tel_radtemp}.

The X-ray observations of PTF\,10tel are presented
and discussed in Section~\ref{sec:X}.

\section{Derivation of mass loss as a function of the H$\alpha$ line luminosity}
\label{sec:Ha}

Here we derive a relation between the mass in
the circumstellar matter (CSM), the H$\alpha$ emission-line luminosity $L_{{\rm H}\alpha}$,
and the radius $r$.
We assume that the CSM has a wind-density profile of the form
$\rho=Kr^{-2}$, where
$K\equiv \dot{M}/(4\pi v_{{\rm w}})$,
$\dot{M}$ is the mass-loss rate,
and $v_{{\rm w}}$ is the wind/outburst velocity.
In this case the particle density profile
is given by (e.g., \erancite{Ofek et al. 2012c})
\begin{eqnarray}
n      & \approx   & \frac{1}{\langle \mu_{{\rm p}}\rangle} \frac{\dot{M}}{4\pi m_{{\rm p}} v_{{\rm w}}r^{2}} \cr
       & \approx     & 3.02\times10^{11} \frac{1}{\langle \mu_{{\rm p}}\rangle} \frac{\dot{M}}{0.1\,{\rm M}_{\odot}\,{\rm yr}^{-1}} \Big(\frac{v_{{\rm w}}}{10\,{\rm km\,s}^{-1}}\Big)^{-1} \Big(\frac{r}{10^{15}\,{\rm cm}}\Big)^{-2}\,{\rm cm}^{-3},
\label{eq:n}
\end{eqnarray}
where
$\langle \mu_{{\rm p}}\rangle$ is the mean number of nucleons per particle
(mean molecular weight).
For our order-of-magnitude calculation, we adopt
$\langle \mu_{{\rm p}}\rangle=0.6$.
Assuming the SN radiation field can ionise
all of the hydrogen in the CSM,
the mass of the hydrogen generating the
H$\alpha$ line is
\begin{equation}
M_{{\rm H}} \approx \frac{m_{{\rm p}} L_{{\rm H}\alpha}}{h\nu_{{\rm H}\alpha}\alpha_{{\rm H}\alpha}^{{\rm eff}} n_{{\rm e}}}.
\label{eq:MH}
\end{equation}
Here $\nu_{{\rm H}\alpha}$ is the H$\alpha$ frequency ($4.57\times10^{14}$\,Hz),
$\alpha_{{\rm H}\alpha}^{{\rm eff}}$ is the H$\alpha$ effective
recombination coefficient,
$\sim8.7\times10^{-14}T_{10{\rm k}}^{-0.89}$\,cm$^{3}$\,s$^{-1}$
(Osterbrock \& Ferland 2006),
where $T_{10{\rm k}}$ is the temperature in units of 10,000\,K.
We assume $T_{10{\rm k}}=1$.
In a wind profile, the integrated mass from radius $r_{0}$ to $r$
is
\begin{equation}
M = \int_{r_{0}}^{r}{4\pi r^{2} Kr^{-2}dr}=4\pi K(r-r_{0}) \approx 4\pi Kr,
\label{eq:Mr}
\end{equation}
where the last step assumes that $r_{0}\ll r$.
By substituting Equation~\ref{eq:n}
into Equation~\ref{eq:MH} and setting it equal to Equation~\ref{eq:Mr},
we can get a relation between the mass-loading parameter $K$,
the H$\alpha$ luminosity, and the radius:
\begin{eqnarray}
K & \gtorder & \Big( \frac{ \langle \mu_{{\rm p}}\rangle m_{{\rm p}}^{2} L_{{\rm H}\alpha} r}{{4\pi h \nu \alpha}^{{\rm eff}}} \Big)^{1/2} \cr
  & \approx    & 7.1\times10^{15} \Big(\frac{L_{{\rm H}\alpha}}{10^{41}\,{\rm erg\,s}^{-1}}\Big)^{1/2} \Big(\frac{r}{10^{15}\,{\rm cm}}\Big)^{1/2}\,{\rm g\,cm}^{-1}.
\label{eq:KLr}
\end{eqnarray}
The reason for the inequality is that in practice, it is possible that not
all of the hydrogen is ionised or that the temperature of the gas
is too high (i.e., $\alpha^{{\rm eff}}_{{\rm H}\alpha}$ depends on temperature).

Assuming the outburst ejected material with a velocity of 2000\,km\,s$^{-1}$
at day $-37$, then at the time the first spectrum was taken (day 6),
the ejecta radius is $r\approx 7\times10^{14}$\,cm.
This radius is consistent with the one
derived from the black-body fit to the spectrum at the first epoch
(see SI Figure~\ref{fig:PTF10tel_radtemp}).
Using the measured flux of the H$\alpha$ broad component
at day 6 ($9\times10^{-15}$\,erg\,cm$^{-2}$\,s$^{-1}$; Table~\ref{tab:spec}),
the luminosity of the H$\alpha$ broad component is $\sim 3\times10^{40}$\,erg\,s$^{-1}$.
Substituting these values into Equation~\ref{eq:KLr},
and assuming a wind velocity
of $2000$\,km\,s$^{-1}$,
we find a mass-loss rate of $\gtorder 10^{-1}$\,M$_{\odot}$\,yr$^{-1}$.
Assuming a one month long outburst this gives a total mass
of about $10^{-2}$\,M$_{\odot}$.
This value is in excellent agreement with the mass-loss rate
we independently find based
on the kinetic-energy arguments and the diffusion time scale.

We note that this derivation assumes that the narrow lines
are due to the interaction of the CSM with the SN radiation field.
One caveat, is that it is possible that some, or all, of the line flux
originates from a shock interaction or
from a slow moving ejecta.
However, the order of magnitude consistency between this mass-loss
estimate and other mass-loss estimators suggest that
such an effect is not significant.

\section{X-ray observations}
\label{sec:X}

The {\it Swift}-XRT observations of SN\,2010mc (PTF\,10tel) in the 0.2--10\,keV
band are listed
in Table~\ref{tab:XRT}.
\begin{table}
\begin{center}
\begin{tabular}{llcc}
\hline\hline
UT Date & Time since disc. & Exp. Time & 2$\sigma$ upper limit \\
        & (day)            & (s)       & (ct\,ks$^{-1}$)        \\
\hline
2010 Aug. 27.0 & 6.8   & 1342 & 2.2  \\
2010 Aug. 31.4 & 11.2  & 2256 & 1.3  \\
2012 July 11.7 & 691.5 & 4818 & 0.62 \\
\hline
\end{tabular}
\caption[Figure~1]{\small {\it Swift}-XRT observations of SN\,2010mc (PTF\,10tel).}   
\label{tab:XRT} 
\end{center}
\end{table}
For each {\em Swift}-XRT image of the SN,
we extracted the number of X-ray counts in the 0.2--10\,keV
band within an aperture of $7.2''$ (3\,pixels) radius
centred on the SN position.
We chose a small aperture in order to minimise
any host-galaxy contamination.
We note that this aperture contains $\sim37$\% of
the source flux (\erancite{Moretti et al. 2004}).
The background count rates were estimated in
annuli around each SN, with an inner (outer) radius of $50''$ ($100''$).
Assuming a photon spectrum $n_{{\rm ph}}\propto E^{-\Gamma}$
with $\Gamma=2$, a Galactic neutral hydrogen column
density of
$2.34\times10^{20}$\,cm$^{-2}$
in the direction of the SN (\erancite{Dickey \& Lockman 1990}),
and correcting for the photometric aperture losses,
we find a 2$\sigma$ upper limit on the luminosity
in the second epoch of $3.5\times10^{41}$\,erg\,s$^{-1}$.

Following \erancite{Immler et al. (2008)},
the X-ray emission from the optically thin
region is given by
\begin{equation}
L_{{\rm X}} \approx \int_{r}^{\infty}{4\pi r^{2} \Lambda(T) n^{2}dr},
\label{eq:Lxn}
\end{equation}
where
$\Lambda(T)$ is the effective cooling function
in the 0.2--10\,keV range.
Assuming an optically thin thermal plasma with a temperature
in the range $10^{6}$--$10^{8}$\,K (\erancite{Raymond et al. 1976}),
we adopt a value of $\Lambda(T)\approx 3\times10^{-23}$\,erg\,cm$^{3}$\,s$^{-1}$.
Substituting Equation~\ref{eq:n} into Equation~\ref{eq:Lxn}
we get
\begin{eqnarray}
L_{{\rm X}} & \approx & 4\pi \Lambda(T) \frac{K^{2}}{\langle \mu_{{\rm p}}\rangle^{2} m_{{\rm p}}^{2}r} \cr
          & \approx   & 2.4\times10^{40} \Big( \frac{\dot{M}}{0.01\,{\rm M}_{\odot}\,{\rm yr}^{-1}} \Big)^{2} \Big( \frac{v_{w}}{2000\,{\rm km\,s}^{-1}} \Big)^{-2} \Big(\frac{r}{10^{15}\,{\rm cm}}\Big)^{-1} {\rm e}^{-(\tau+\tau_{{\rm bf}})}\,{\rm erg\,s}^{-1}.
\label{Lx}
\end{eqnarray}
Here, we added a correction factor due to absorption in the wind,
where $\tau$ and $\tau_{{\rm bf}}$ are the Thomson and bound-free optical depths, respectively.
In our case the Thomson optical depth will be 1 for $\dot{M}\approx0.1$\,M$_{\odot}$\,yr$^{-1}$
(e.g., Ofek et al. 2012c).
Although the Thomson optical depth is well known,
when the optical depth is of the order of a few,
Compton scattering is expected to reprocess more energetic
photons into the 0.2--10\,keV band (Chevalier \& Irwin 2012; Svirski et al. 2012).
Since the exact X-ray spectrum is not known
(Katz et al. 2011; Svirski et al. 2012), a proper calculation of
$L_{{\rm X}}$ when $\tau\gtorder1$ is not straightforward.
Regarding the bound-free optical depth,
although it is likely that all the hydrogen is ionised,
Chevalier \& Irwin (2012)
argued that if the shock velocity is below $\sim 10^{4}$\,km\,s$^{-1}$
some of the metals will be neutral.
In this case bound-free absorption will be important
and it is given by (Ofek et al. 2012c)
\begin{equation}
\tau_{{\rm bf}} \approx 15 \Big(\frac{\dot{M}}{0.01\,{\rm M}_{\odot}\,{\rm yr}^{-1}}\Big) \Big(\frac{v_{{\rm w}}}{2000\,{\rm km\,s}^{-1}}\Big)^{-1} \Big(\frac{r}{10^{15}\,{\rm cm}}\Big)^{-1} \Big(\frac{E_{{\rm X}}}{1\,{\rm keV}}\Big)^{-2.5},
\label{tau_bf}
\end{equation}
where $E_{{\rm X}}$ is the X-ray energy.

If we ignore the Thomson optical depth (i.e., $\tau=0$), and if we
assume a flat spectrum (i.e., constant number of photons per unit energy)
and the {\it Swift}-XRT spectral response,
from Equations~\ref{Lx} and \ref{tau_bf} we find 
that the mass-loss rate is either smaller
than about 0.1\,M$_{\odot}$\,yr$^{-1}$
or larger than about 1\,M$_{\odot}$\,yr$^{-1}$.
The low bound of $\dot{M}\ltorder 0.1$\,M$_{\odot}$\,yr$^{-1}$
is consistent with the other mass-loss estimates.

\section{Radio observations}
\label{sec:Radio}

We observed SN\,2010mc (PTF\,10tel) with the Jansky Very Large Array
(JVLA\footnote{The Jansky Very
Large Array is operated by the National Radio Astronomy Observatory (NRAO),
a facility of the National Science Foundation operated under
cooperative agreement by Associated Universities, Inc.}) on 2012 Sep. 10,
about two years after the SN discovery.
The observation was undertaken in the K band (21.8\,GHz). We used J1727$+$4530 for phase
calibration and 3C\,286 for flux calibration.
The data were reduced using the AIPS software. We did not detect SN\,2010mc with an upper
limit of 0.2\,mJy (3$\sigma$).

\section{Diffusion time scale limit on mass loss}
\label{sec:Tdiff}

Another observable that can be used to constrain
the mass in the CSM is the rise time of the SN light curve.
If a considerable amount of material is present between the SN
and the observer,
then photon diffusion will slow down the rise time of the SN light curve.
Therefore, the maximum observed SN rise time can be used to put an upper
limit on the amount of mass between the SN and the observer.
The diffusion time scale in a wind profile is given by
(e.g., \erancite{Ginzburg et al. 2012})
%
\begin{eqnarray}
t_{{\rm diff}} & \approx &\frac{\kappa K}{c}[\ln{\Big(\frac{c}{v_{{\rm sh}}}\Big)}-1] \cr
             & \approx   & 0.08
                         \Big(\frac{\kappa}{0.34\,{\rm cm}^{2}\,{\rm gr}^{-1}}\Big)
                         \Big(\frac{\dot{M}}{10^{-2}\,{\rm M}_{\odot}\,{\rm yr}^{-1}}\Big)
                         \Big(\frac{v_{{\rm w}}}{2000}\Big)^{-1}
                         \Big[\ln\Big(30\frac{v_{{\rm ej}}}{10^{4}\,{\rm km\,s}^{-1}}\Big)-1\Big]\,{\rm day.}
\label{tdiff}
\end{eqnarray}
Here $v_{{\rm ej}}$ is the SN shock velocity.
In our case, assuming the rise time of the SN is about seven days, we can use
Equation~\ref{tdiff} to set an upper limit, $\dot{M}<0.4$\,M$_{\odot}$\,yr$^{-1}$.

\section{Alternative interpretations}
\label{Interp}

Here we consider alternative interpretations regarding
the nature of the precursor bump and show that
they are not consistent with the observations.
One possibility is that this feature is due to
a shock breakout (e.g., \erancite{Colgate 1974})
prior to the SN rise.
However, given the feature's luminosity, 
its duration is at least an order of magnitude too long
to be consistent with a shock breakout
(e.g., \erancite{Nakar \& Sari 2010};
\erancite{Rabinak \& Waxman 2011}).

\begin{figure}
\centerline{\psfig{file=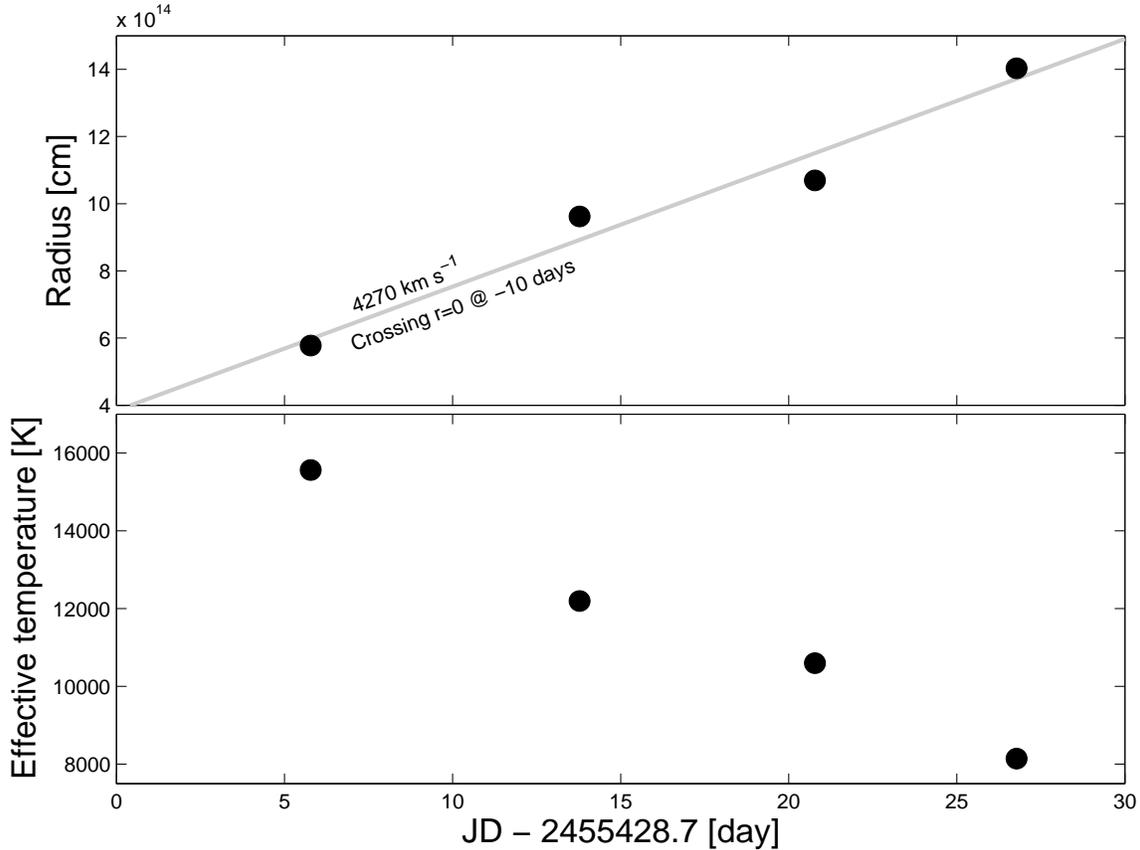,width=6in,angle=0}}
\caption[]{\small 
Black-body temperature and radius derived from the spectra of SN\,2010mc (PTF\,10tel).
The temperature estimate should be regarded as a lower limit,
because the peak of the emission is outside the optical region
and metal line blanketing can reduce the flux in the blue parts
of the spectrum.
The reduction in temperature is consistent with the fact that the He\,I lines
are becoming weaker at later epochs (SI Figure~\ref{fig:PTF10tel_spec_Halpha}).
We note that the best-fit temperature at day 6 (near the first spectroscopic
observation) based on the broad-band {\it Swift}/UVOT and P60 photometry is about 19,000\,K.
The nondetection of He\,II lines also suggests that the temperature
is not much higher than the value derived here.}
\label{fig:PTF10tel_radtemp} 
\end{figure}

A second possibility is that the first bump represents a SN explosion
while the second peak is due to the interaction of the SN ejecta
with dense CSM
and the conversion of the SN kinetic energy into radiated luminosity.
However, this scenario can be rejected based on the fact
that 
by fitting a black-body spectrum to the spectra
(see SI Figure~\ref{fig:PTF10tel_radtemp}),
we can find the temperature and photospheric radius
evolution between days 5 and 27.
These fits suggest that the photosphere is expanding with a mean
velocity of $\gtorder$4300\,km\,s$^{-1}$, and that the black-body
radius at day 6 is $\sim6\times10^{14}$\,cm.
Extrapolating backward, this suggests that the SN outburst
happened after day $-10$.

Another argument against this scenario is that
the SN spectrum at day $27$ after discovery
(main text Figure~2), 
shows a P-Cygni profile with
a velocity of $\sim10^{4}$\,km\,s$^{-1}$.
At the same time, the black-body radius of the SN photosphere
is $\sim1.5\times10^{15}$\,cm.
These values are roughly consistent (to within 30\%)
with an explosion time at day 0, and are less consistent
with explosion at day $-37$\,day relative to discovery.
A caveat in this statement is that
the photospheric radius which controls
the black-body temperature of the continuum is not necessarily
at the same position where the P-Cygni absorption is generated.

\section{The Type Ibn SN\,2006jc}
\label{sec:2006jc}

Another possibly related event is SN\,2006jc (\erancite{Nakano et al. 2006}).
Unlike SN\,2010mc (PTF\,10tel) and SN\,2009ip,
this was a hydrogen-stripped Type Ibn SN.
In this case
an optical outburst was observed in 2004,
about two years prior to the SN explosion (\erancite{Nakano et al. 2006}),
creating a dense circumstellar shell
(\erancite{Foley et al. 2007};
\erancite{Immler et al. 2008};
\erancite{Smith et al. 2008};
\erancite{Ofek et al. 2012c}).
\erancite{Foley et al. (2007)} suggested that 
the progenitor was a Wolf-Rayet star that had an outburst two years 
prior to its explosion, creating the CSM; it may
have recently transitioned from the LBV phase.
However, \erancite{Pastorello et al. (2007)} argued that 
SN\,2006jc took place in a binary system
in which one star (a Wolf-Rayet star)
was responsible for the SN explosion
while the companion was a
luminous blue variable (LBV) that generated
the eruption in 2004.
Our argument regarding the causality between
pre-explosion outbursts and the SN explosions
disfavours the explanation
given by Pastorello et al. (2007) for SN\,2006jc.
However, for SN\,2006jc, the time difference between the 
precursor event and the explosion is about two years
(rather than one month),
rendering the argument a 
factor $\sim20$ weaker than in the case of PTF\,10tel.

\section{The probability of detecting a precursor outburst prior to the explosion}
\label{sec:precprob}

The specific precursor event reported in this paper
was found as a part of a search for outbursts among type-IIn SN
in PTF data.
The sample we used included only four nearby type-IIn SNe with good
pre-explosion coverage.
Nevertheless, in estimating the probability
of detecting a precursor prior to the explosion
we assumed a sample size of 20 (five times higher than our actual sample size),
accounting for other samples in which outburst events could have been detected
(see details below).
This sample size thus facilitates a conservative
rate estimate.

Thus far the brightest precursor outbursts (e.g., PTF\,10tel)
have an absolute magnitude of about $-16$.
Using previous-generation transient surveys, such bursts can be
detected up to a distance of $\sim100$\,Mpc,
while the new-generation surveys (e.g., PTF) may be able to detect
them to somewhat larger distances.
Among the 165 SNe~IIn listed in the IAUC website,
about 70 have distances below 100\,Mpc.
Many of these SNe were found using small telescopes
which are limited by their depth.
Based on the PTF experience, we estimate that
not more than 20\% of these SNe have sufficiently good pre-explosion
observations that will allow the detection of a precursor
prior to the explosion.
In addition, PTF has already found 91 SNe~IIn (some
listed in the IAUC website),
of which only a few are sufficiently nearby, and have high cadence
and almost continuous observations prior to the explosion.
Thus, we conclude that up to 20 SNe~IIn
discovered so far have good enough pre-explosion
observations to find precursor outbursts.

We note that even if we assume that {\it all} of the $\sim 250$
SNe~IIn discovered so far have sufficiently good observations
prior to the SN explosion, this will not change
our conclusion that precursor
outbursts are more common, or are causally connected with
the final stages of massive-star evolution.

Finally, we note that we cannot completely rule
out the possibility that the SN and precursor
are two unrelated events (e.g., explosions of two different stars in
a star cluster).
However, we think this is an unlikely possibility
given the evidences presented in this paper,
including consistency of the mass-loss estimators
with the properties of the outburst (luminosity and time before SN),
and the small number of type-IIn SNe in which we searched for
precursor events.

\section{Theoretical determination of the precursor mass loss}
\label{sec:MassLossTheory}

The precursor is characterised by a rapid brightening of the radiative flux which is
sustained at a highly super-Eddington level for a month. The relatively short time
scales imply that some efficient mechanism exists to very quickly transport the flux out,
even faster than convection can. The mechanism suggested by
\erancite{Quataert \& Shiode (2012)},
where the energy is transported outward by acoustic waves, offers a natural explanation.
However, this mechanism can operate only in regions with a sufficiently
high density.
At a region where the density, $\rho$, satisfies
$L \approx L_{{\rm conv,max}} \approx 4 \pi r^2 \rho v_{{\rm s}}^3$ (where
$v_{{\rm s}}$ is the speed of sound),
the acoustic energy flux $L$ is converted into a radiative flux $L$.
The mass-loss rate expected in such a case is
$\dot{M} \approx 4 \pi r^2 \rho v_{{\rm s}} \approx L/v_{{\rm s}}^2$.
Assuming $v_{{\rm s}} \approx 60$\,km\,s$^{-1}$
at the base of the wind gives roughly
8\,M$_{\odot}$ for the mass ejected during the precursor.
A similar problem exists for the great (super-Eddington) eruption of
$\eta$~Carina. The naive expectation for the mass loss is orders of
magnitude larger than what was actually observed (\erancite{Shaviv 2000}). 

To overcome this discrepancy, it was suggested that instabilities reduce the
effective opacity of super-Eddington atmospheres. This implies that a wind
is accelerated only from lower-density regions where a typical scale height
in the atmosphere becomes of order unity and the effective opacity returns
to its microscopic value. Under these assumptions, the mass-loss was
determined by Shaviv (2001) to be
$\Delta{M} = W \Delta{T} (L-L_{{\rm Edd}}) / (c v_{{\rm s}})$,
where $\Delta{T}$ is the duration of the outburst and $L$ is the 
luminosity at the base of the wind
(where part of the latter is converted into mechanical energy of the mass loss).
$W$ is a dimensionless constant which depends on the geometry of the
instabilities, and empirically was found to be around 3 ($\pm 0.5$ dex).

We can use this expression to predict the expected mass loss from the
super-Eddington precursor. To do so we have to add the mechanical energy
of the wind to the observed luminosity and obtain the total luminosity at the
base of the wind. The mechanical energy itself is given by
$L_{{\rm mech}} = \Delta{M}/2 (v_{\infty}^2 + v_{{\rm esc}}^2)$.
Here $v_{\infty}$ is the terminal velocity at infinity
and $v_{{\rm esc}}$ is the escape velocity.
We crudely take
$v_{{\rm esc}} \approx v_{\infty}$. Since the luminosity at the base is
significantly super-Eddington, we can ignore the $L_{{\rm Edd}}$ term.
If we further consider that $v_{{\rm s}} \approx 30$\,km\,s$^{-1}$, we obtain
$\Delta{M} \approx 0.05$\,M${_\odot}$ (with an uncertainty of about 0.5 dex).
This value sits comfortably in the observationally determined
range of possible mass loss in SN\,2010mc (PTF\,10tel).

\clearpage

\bibliographystyle{nature-pap}

\end{document}